# Simultaneous Energy Harvesting and Hand Gesture Recognition in Large Area Monolithic Dye-Sensitized Solar Cells

*Gethin Thomas, Adam Pockett, Kris Seunarine, and Matt Carnie\**

Materials Science and Engineering & SPECIFIC-IKC, Faculty of Science and Engineering, Swansea University, Swansea, UK

m.j.carnie@swansea.ac.uk

**ABSTRACT**

Internet of Things (IoT) devices have become prevalent, embedding intelligence into our environment. It is projected that over 75 billion IoT devices will be connected by 2025 worldwide, with the majority being operated indoors. Dye-sensitized solar cells (DSSC) have recently been optimized for ambient light, having the capabilities of providing sufficient energy for self-powered IoT devices. Interaction with digital technologies, termed Human Computer Interaction (HCI), is often achieved via physical mechanisms (e.g. remote controls, cell phones) which can hinder the natural interface between users and IoT devices, a key consideration for HCI. What if the solar cell that is powering the IoT device can also recognize hand gestures which would allow the user to naturally interact with the system? Previous attempts to achieve this have necessarily employed an array of solar cell/photodiodes to detect directionality. In this work, we demonstrate that by monitoring the photocurrent output of an asymmetrically patterned monolithic (i.e., single cell) DSSC, and using machine learning, we can recognize simple hand gestures, achieving an accuracy prediction of 97.71%. This work shows that, DSSCs are the perfect choice for self-powered interactive technologies, both in terms of powering IoT devices in ambient light conditions and having aesthetic qualities that are prioritized by users.  As well as powering interactive technologies, they can also provide a means of interactive control.

## 1. INTRODUCTION

There has been unprecedented rise in the number of IoT devices all over the world, with a projected 75 billion connected devices by 2025 [1]. The IoT ecosystem is expected to grow to an incredible one trillion by 2035 [2], with a substantial amount placed in indoor environments, being used in a wide range of applications, having the potential to cover the social, environmental, and economic impacts.

Due to the large increase in IoT nodes (being the key parameters of collecting, processing, and transmitting the relevant data between each node, and to the internet), it has resulted in security issues [3] due to the limited memory, processing capabilities and heterogeneity, with autonomous operation for powering IoT nodes having the biggest critical issues that is preventing the growth of indoor IoT ecosystem [4]. A promising solution in providing sufficient power in indoor conditions is to harvest indoor energy through ambient lighting. This would result in prolonging the lifespan of these devices, key in reducing the complexities of replacing batteries in inaccessible areas, having a greater economic appeal and technical viability for security. This could either be to aid the battery life of the IoT nodes, storing the energy harvested, or preferably work battery-less [5].

As computers became widespread, the field of Human Computer Interaction (HCI) - studying the interaction between users and computers – has become widespread. One of the primary drivers in technology for HCI is the commitment to value human activity and experience. This paper focuses on the physical connection between the user and the self-powered IoT node, achieving a suitable and easy-to-use user interface (UI). Most modern applications involve interaction through a touch screen or physical mechanisms (such as remote controls, cell phones), allowing the user to directly interact with the display. However, this can hinder the natural interface between users and IoT devices, a key consideration for HCI.

Hand gesture recognition in IoT applications offers a compelling and intuitive way to interact with and control a wide range of connected devices. It can improve the users experience, mimicking real-world physical actions, making them easy to understand and accessible to people of various age groups and backgrounds. This includes individuals with physical disabilities, making it possible for them to control IoT devices and access technology that might otherwise be challenging. Hand gesture also allows the user to control IoT devices without physical contact, giving them the convenience of hands-free (such as if their hands are occupied or dirty), but also reduces the risk of contamination, which can be especially important in environments like healthcare, smart homes, and public spaces [6].

Conventional real-time hand gesture recognition systems include visual sensor based techniques [7], using captured videos from a camera to identify the gesture using a computer algorithm. Other techniques involve the user wearing a data glove [8], having integrated sensors embedded to transform the hand motion information into electrical signals. Both of these techniques require a large amount of input power, including limitations with latency issues and large data outputs [9][10].

By integrating an energy harvesting module to harvest energy but to also act as a sensor to detect the user's gesture/interaction with the IoT node, would be an advantageous system to lower the overall energy consumption of the IoT node compared to other gesture detection techniques [11][12][13]. There are two different types of hand gestures, being static and dynamic gestures. Static involves the motion of different hand shapes, whilst dynamic gestures are generally described according to hand movements. This paper concentrates on dynamic gestures, as motion of hand is a more natural response when operating a smart device integrated with a display.

Directional based gesture detection coinciding with energy harvesting has been achieved by attaching several photodiodes to the system. As the system knows the location of each photodiode; when a hand gestures over the device, the cast in shadow causes the photocurrent to be lowered. The pattern in photocurrent reduction (for each photodiode) can then be analysed to determine the hand direction [10][13][14][15]. Whilst this does allow efficient and reliable hand gesture recognition, due to each photodiode working independently, the set-up requires individual wiring to the system, resulting in a complex assembly and not as visually appealing UI. Ma et al. [11] in 2019, demonstrated the feasibility of hand gesture recognition system from a photocurrent signal response of a monolithic opaque silicon cell, and two different transparencies (20.2% and 35.3%) of an organic cell (PBDB-T: ITIC). This achieved a hand gesture recognition accuracy of 96%, by including signal pre-processing techniques, consisting of dynamic time warping (DTW) and Z-score transformation. The cleaned signal is then classified, from an extensive dataset, using machine learning (ML) techniques. Whilst it seems that SolarGest has achieved a monolithic photovoltaic (PV) hand gesture recognition, there are some limitations with the device: Six different gestures were achieved; however, not directional based gestures. This means that this device cannot differentiate if the user is swiping right or swiping left over the cell due to the even layer of semiconductor material on the substrate, providing a constant photocurrent signal. Furthermore, the pre-processing technique restricts further analysis of the device. Following the use of DTW, the author does this by using two signals of the same gesture. However, this cannot be achieved if the gesture is done during live gesture recognition as the device cannot predict before what motion of the hand is accomplished.

Previous attempts to design an integrated PV hand gesture recognition device have involved either a complicated set-up with multiple PV cells, or gestures that are not fully natural and comfortable to accomplish. Therefore, the goal in this paper was to build a prototype, involving a monolithic dye sensitised solar cell (DSSC), that can distinguish from basic directional hand gesture swipes (right, left, up, down) using the photocurrent output from the PV. Both DSSCs and organic photovoltaics (OPVs) were initially considered, having many similar characteristics other than its high efficiencies in indoor light, including low cost, non-toxic material, colourful, transparency, and their flexibility [16][17]. However, DSSCs have been shown to have a simpler and easier fabrication method (being a very attractive and appealing device during IoT node design), especially based on evidence for larger active area solar cells, such as the 20.25 cm$^2$ DSSC, showing good performance in indoor conditions [18].

## 2. EXPERIMENTAL METHODS

Unless otherwise stated, DSSCs are produced on fluorine doped tin oxide (FTO) coated glass (TEC 15, 2.2 mm) substrates from XOP Glass. For blocking layer (BL) deposition, a solution of titanium diisopropoxide bis(acetylacetonate) is diluted with IPA, in a 1:9 volume ratio. The BL is deposited via spraying at 200°C followed by heating to 500°C. The TiO$_2$ working electrode (WE) is made by screen-printing Solaronix Ti-Nanoxide T/SP paste. The screens used are manufactured by MCI Precision Screens, (product code 245 43-80 and mesh size 45°, 13m). The screen-printed paste is then sintered for 30 minutes at 500°C. The dye used was Ruthenizer 535-bisTBA (commonly known as N719) from Solaronix. It was dissolved in ethanol (0.5 mmol). The TiO$_2$ working electrodes were left to dye overnight. The counter electrode (CE) is formed by thermally decomposing 5 mM of chloroplatinic acid in IPA at 550°C. A 25 μm thermoplastic gasket (Solaronix -Meltonix) is used to seal the WE to the CE.

Two different types of electrolytes are used throughout this paper. Initially, the commercial idolyte AN-50 (Solaronix) is used for the initial flower pattern design. For the computational optimised patterns, the AN-50 electrlolyte became unavailable so, a homemade (HM) solution was used instead, consisting of 0.6 M 1-methyl-3-n-propylimidazolium iodide, 0.1 M lithium iodide, 50 mM iodine, and 0.5 M 4-tert-butylpyridine.

An enclosed low light simulator was used to measure the performance of DSSCs in indoor conditions, by providing a diffused LED light source with a colour temperature of 2700 K. IV curves are measured with a Keithley SMU. The LED array is calibrated using an Ocean Insight spectrophotometer, which is set-up to measure the photometry in lux values through the 'OceanView' software. The photocurrent gesture response is captured using a transimpedance amplifier TZA500 and data acquisition device (DAQ) (NI USB-6210), converting the analogue electrical signal to digital, through a 16-bit resolution. An LED array sat above the DSSC is used as a consistent light source when collecting the gesture photocurrent signal. We created an application through LabVIEW that enables the user to capture a 2 s gesture multiple times, when the photocurrent goes below a set threshold. This allows the ease of multiple gesture measurements without the need of resetting the device per gesture. After completing the number of gestures required for collection, the data can be all saved under one Excel document for further analysis using MATLAB.

## 3. RESULTS AND DISCUSSION

We initially hypothesised that a monolithic DSSC with an asymmetrically sized active area pattern could be used to detect directional hand gestures by monitoring the photocurrent signal when the user's hand was passed over it. To test this hypothesis, an 8 x 8 $cm^2$ monolithic DSSC was fabricated, and covered with cardboard featuring an asymmetrical cut-out pattern - see Figure 1(a). The photocurrent response was then measured for four directions: right, left, up, and down, by swiping a hand close to the face of the DSSC.

Figure 1(b) shows some promising differences between each directional photocurrent signal. The next step was to see if an asymmetrically patterned $TiO_2$ could give similar results. We chose an aesthetically pleasing flower pattern shown in Figure 1(c).

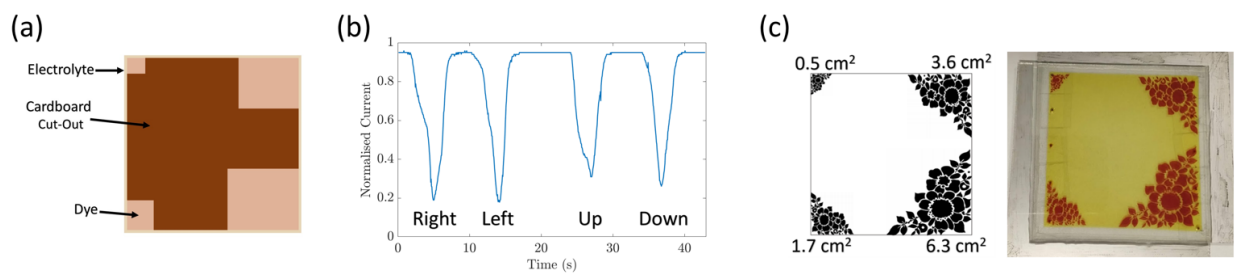

*Figure 1: (a) Schematic of initial cut-out cardboard hypothesis. (b) Photocurrent response using the cut-out cardboard. (c) Asymmetrical pattern design.*

Following the gesture data collection method, the signal output of 4 hand gestures was collected. The photocurrent signal of each gesture shows some slight difference in the shape, seen in figure 2. One hundred and twenty samples of each gesture were collected under light intensities of approximately 1000 and 2500 lux. This was controlled by continuously placing a VEML7700 lux meter connected to a display. The light source can then be adjusted to the desired intensity using a power supply. The total dataset consisted of 960 gestures for analysis.

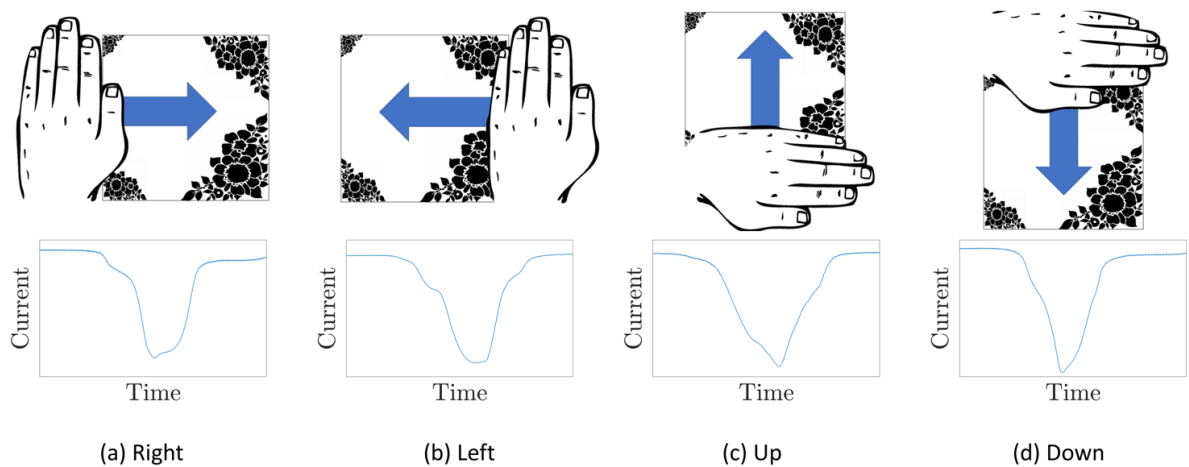

*Figure 2: Sketches of the 4 hand gestures performed over the DSSC. The second row shows the photocurrent response for each gesture at 1000 lux.*

Our first approach to signal analysis was to use thresholding, which relies on extracting specific features that can differentiate each gesture photocurrent signal. The process manually goes through the 'if' loop process, using pre-defined values that have been selected at the start; and thereby separating each gesture. The technique extracted the normalised current values when the gradient was zero for each gesture, however, this feature was only amplified when the light intensity was higher than 2500 lux. Due to this, the gesture prediction accuracy severely decreases below 2500 lux (see supplementary information, Figures S1-2).

With the threshold technique being unfavourable, an alternative method, which was also used by Ma et al. [11], is machine learning (ML). Supervised classification identifies which category (in this case, the four gesture sweeps) a new set of data belongs to, predicted from a training model built from another set of data that has already been categorised [19]. The general classification framework involves data pre-processing the photocurrent time series, extracting specific features, and finally training a classifier. MATLAB was used as the processing software for building the training model and executing the trained model against a new set of data.

The data pre-processing technique closely follows the successful method applied by Ma et al. [11], which can mitigate some parameters that can affect the photocurrent signal of the same gesture, including quality of DSSC fabrication (efficiency and fill factor); environmental conditions (light intensity); and user's parameters (height from DSSC, speed of gesture, hand angle and size). Initially, Z-score transformation is used to align the signal amplitudes, with a mean of 0 and standard deviation of 1, as seen in Figure 3a. Dynamic time warping (DTW) is then used to solve the temporal misalignment. It does this by distorting the duration of both data sets, to make them appear on a common time axis at the same location. The default distance metric is used, specified as 'Euclidean' which is the root sum of squared differences. However, as previously mentioned, Ma et al. [11] uses two signals of the same gesture, which prevents any live data to be predicted. To overcome this, a reference curve is used instead, which dynamic time warps the measured photocurrent signal (any of the four directional hand gestures) to the reference curve, shown in Figure 3b. The function 'alignsignals' in MATLAB is also used before and after DTW which improves the alignment of each gesture shown in Figure 3c.

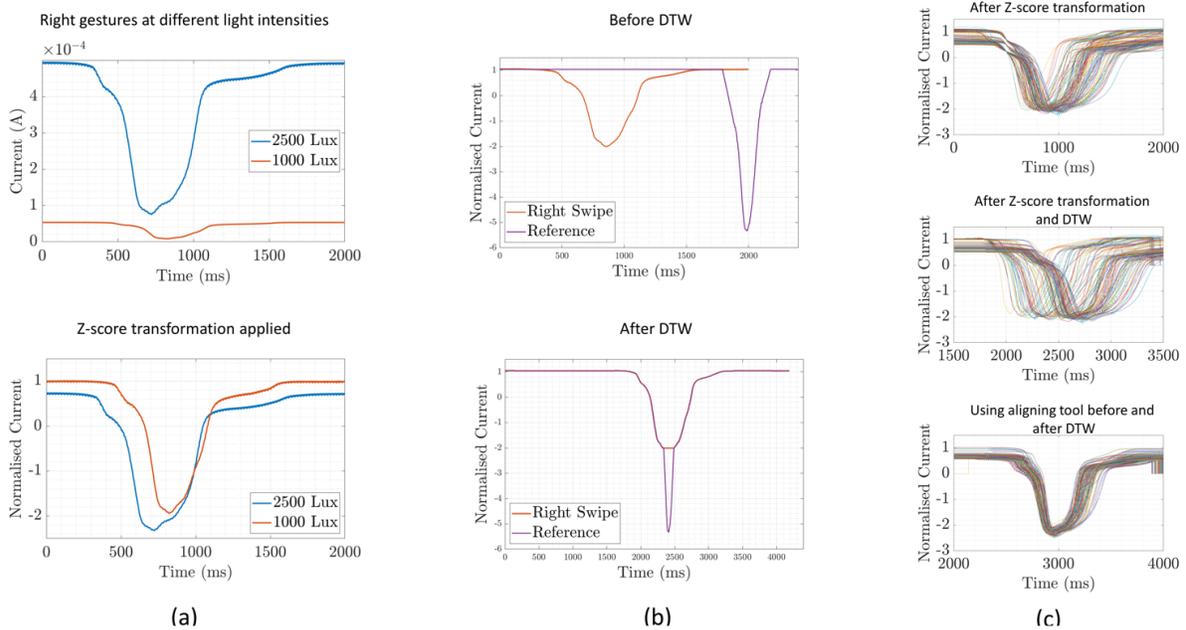

*Figure 3: (a) Amplitude alignment using Z-score transformation. (b) DTW technique using a reference curve. (c) Example of large dataset using DTW and alignment tool.*

After pre-processing the data, feature selection extracts specific values that reduce the number of values of each gesture. By selecting a small subset from the original data, it tends to lead towards an increased trained model performance, better data interpretability and lowering the computational cost [19]. Several statistical features were initially extracted, however, after undergoing machine learning to train the models, the overall outcome was unsuccessful. An alternative feature extraction method also used by Ma et al. [11] was discrete wavelet transform (DWT) coefficients, which decomposes a signal at various resolutions and different frequency domain, capturing both temporal and frequency information.

The detail coefficients are extracted on the selected coefficient level used. Ma et al. [11] does this by halving the sampling rate, to obtain the highest frequency contained in the signal, following

Nyquist Theorem. In this case the sampling rate is 1kHz, therefore 500 Hz is the contained highest frequency. The fast Fourier transform (FFT) has a frequency gesture that lies below 5 Hz, as seen in Figure 4a. Therefore, the level 6 DWT is required, as it covers the gesture frequency, as the range is [0, $500/2^6$] Hz = [0, 7.8] Hz. Level 5 detail coefficients are also collected, to compare if a larger frequency range is beneficial or not for machine learning. Figure 4b shows the DWT process, halving the amount of detail coefficient points as the wavelet level increase from 5 to 6, also showing the approximation coefficients at level 6. The example extracts the detail coefficients of a single right gesture through Daubechies2 (db2) wavelet.

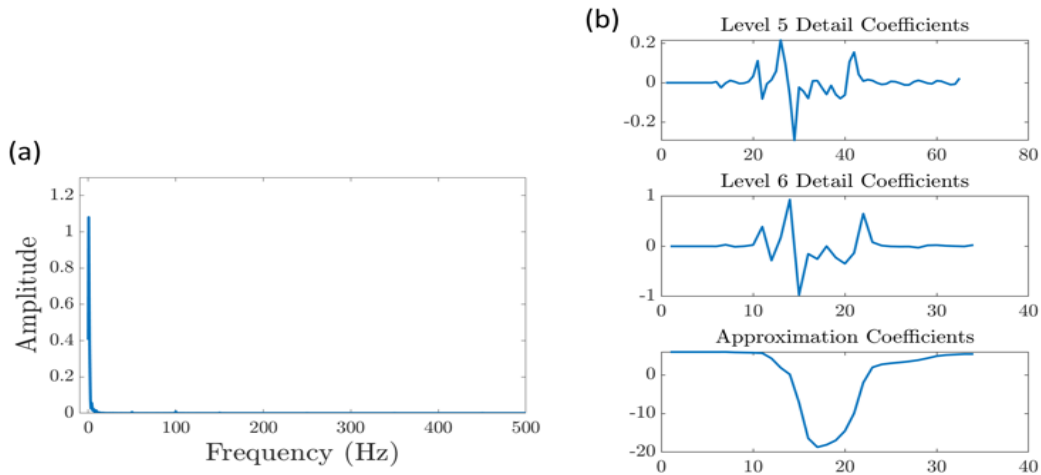

*Figure 4: (a) FFT analysis of a single photocurrent signal. (b) DWT detail coefficients up to level 6 with approximation coefficients level 6.*

There are several other different types of wavelets available to use, with MATLAB capable of using these wavelet families: Daubechies, Coiflets, Symlets, Fej´er-Korovkin, Discrete Meyer, Biorthogonal, and Reverse Biorthogonal. Ma et al. [11] undertook a wavelet analysis comparing five different wavelets: Haar(haar), Daubechies1(db1), Daubechies2(db2), Daubechies4(db4), and Coiflet2(coif2), with db2 having a slightly higher recognition accuracy. These five different wavelets are also investigated during machine learning in this paper.

During the set-up of machine learning, half of the data from the 1000 lux and 2500 lux (done randomly but an equal amount for each gesture and light intensity) are placed in one file for classification; totalling 480 photocurrent signals (120 for each gesture). The data follows the pre-processing technique, involving Z-transformation and DTW (including the alignment function), leading to level 5 and 6 detail coefficients being extracted. After this, the machine learning classifiers are trained for gesture recognition. This is done through the MATLAB application called classification learner. As there is uncertainty as to which classifier model is best, all models capable on MATLAB are trained to find the best accuracy. A 10-fold cross-validation is selected, which involves dividing the data randomly into equally sized folds (k), where k is defined by the user. One of the folds is used as the test set with all the other folds being used as the training set.

This is repeated for each unique fold, averaging the observed errors to form the k-fold estimate. To additionally validate the trained model, an unseen dataset (data in which the trained model has not seen) consisting of another 480 gestures (being the other half of the gestures not used during classification, at 1000 and 2500 lux) was used to test the trained model. Each trained model predicts the gesture for each unseen photocurrent signal, with the accuracy of success being recorded.

The best model from the flower pattern design after evaluation was found to be the level 5, wavelet feature Db1, under the classifier KNN (weighted), achieving 97.71%. The confusion matrix of the model is shown in Table 1 below.

Table 1. Confusion Matrix of best evaluated model from flower pattern DSSC

| True Class | Right | Left | Up | Down |
|---|---|---|---|---|
| Right | 115 | 2 | 0 | 3 |
| Left | 0 | 118 | 2 | 0 |
| Up | 1 | 1 | 116 | 2 |
| Down | 0 | 0 | 0 | 120 |

Predicted Class

After we successfully detected four gestures on a flower pattern active area, an alternative approach we hypothesised was etching the FTO layer from the working electrode, whilst screen printing a uniform $TiO_2$ layer. This approach would result in higher output from the cell, whereby the entire area is power generating and also gives the appearance of a normal DSSC, but in this case, the asymmetrical etched FTO on the working electrode allowing for a distinguishable difference in the photocurrent output for each of the four directional hand gestures. The FTO was etched by using zinc oxide and 4 M of hydrochloric acid (HCL), scrubbing the layer away before applying the BL. Following on from previous results, Haar, Db1 and Db2 wavelet features of level 5 and 6 detail coefficients were selected, with the SVM, KNN and Ensemble trained model classifiers being collected when testing the etched FTO DSSC, due to high accuracies from previous results. After evaluation, the etched FTO achieved a remarkable prediction accuracy of 99.79%, with the confusion matrix from the best trained model of SVM Quadratic, at a DWT of level 6 shown in Table 2.

Table 2. Confusion Matrix of best evaluated model from etched FTO DDSC

| True Class | Right | Left | Up | Down |
|---|---|---|---|---|
| Right | 119 | 1 | 0 | 0 |
| Left | 0 | 120 | 0 | 0 |
| Up | 0 | 0 | 120 | 0 |
| Down | 0 | 0 | 0 | 120 |

Predicted Class

Additional gestures alongside the original four gestures can be tested and analysed, finding the ML capabilities and limits. Four diagonal directional gestures were added to the flower pattern

DSSC, as seen in Figure 5, with the photocurrent response shown on the second row. Initial comparison between the four diagonal gestures shows similarities in the shape of the photocurrent signal, with a resultant evaluated gesture prediction of 60.21% for the eight directional gestures.

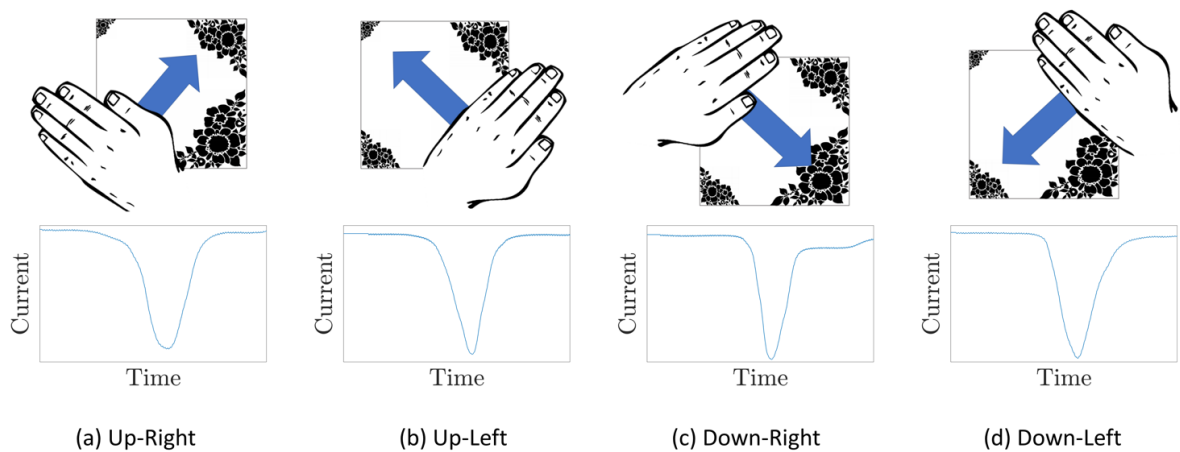

*Figure 5: Sketches of the 4 additional hand gestures performed over the DSSC. The second row shows the photocurrent response for each gesture at 1000 lux.*

Expanding the number of gestures recognised would enhance the interaction, allowing users to interact with devices in a more nuanced and expressive manner. It would also provide a richer user experience and increased functionality, by enabling the device to perform a wider array of functions. Overall, by increasing the number of recognised hand gestures by a DSSC offers a more versatile, user-friendly, and adaptable gesture recognition experience.

To improve the prediction accuracy of detecting eight directional gestures, we created a model on LabVIEW which outputs the photocurrent signals of each gesture swipe using different active area patterns designed for optimisation. There are some factors which are not considered in the model, these include: light intensity; and the user's parameters (such as the speed and height of the hand), except for hand size. The hand size parameter can be controlled by altering the width size of the scanning area bar. Each design was created in bitmap form, through the design software CorelDRAW, of 800 x 800 black (representing the active area) and white (representing no active area) pixels, with the output signals exported from the LabView Model for each of the eight gestures. Some initial ideas for new patterns revolved around the preliminary design (flower pattern), which achieved a remarkable 97.71% for the four directional gestures. Other new designs were created randomly with a mixture of different shapes to explore the output photocurrent signal. Every design created and its computed photocurrent response are stored in the supplementary information, in Figures S3-45.

It was seen that the largest difference in photocurrent signal between each directional gestures was achieved through pattern 1, in Figure 6a. Two additional adaptions (patterns 2 and 3) are also investigated seen in Figure 6b and c, both slightly increasing the overall active area for optimised power output. Following the same machine learning procedure as before, pattern 1,2 and 3 achieved the highest evaluated accuracy of 62.08%, 57.40% and 83.23% respectively, with the confusion matrix of the best evaluated model from pattern 3 shown in Table 3. The best computed design was from pattern 1, thus it was unexpected to see pattern 3 achieving better gesture prediction accuracy.

The performance characteristics for each pattern, which also compares the initial flower pattern and etched FTO are shown in Figure 6d, e, f, and g. Figure 6d shows a significant drop in open-circuit voltage ($V_{OC}$) between 5000 and 100 lux for pattern 1 and 2. This is the probable cause for the poor accuracy predictions from the design patterns 1 and 2, only achieving 60.21% and 57.40%, as the data capturing method involves measuring the voltage difference between a known resistor, in order to measure the current. If there is a lower voltage output, then the differential before and after will be lower, reducing the photocurrent output accuracy and hence reduce the amplification of the shape of the curve. As previously mentioned, due to issues with ordering the AN-50 electrolyte, a homemade electrolyte solution involving 50 mM of iodine was instead used for the optimised active area patterns. Further tests, supplied in the supplementary information, in Figures S46-47, led to the discovery that increasing the horizontal distances between active areas results in significantly decreased $V_{OC}$ and fill factor, due to charge recombination which is known to be faster using the homemade electrolyte than the AN-50 from Solaronix. We believe pattern three performed better, due to the screen-printed dots reducing horizontal distance between active areas. The performance characteristics also shows that the etched FTO provides the highest output power compared to the other DSSC design iterations which is to be expected as the geometric fill factor is close to 100 % in this case.

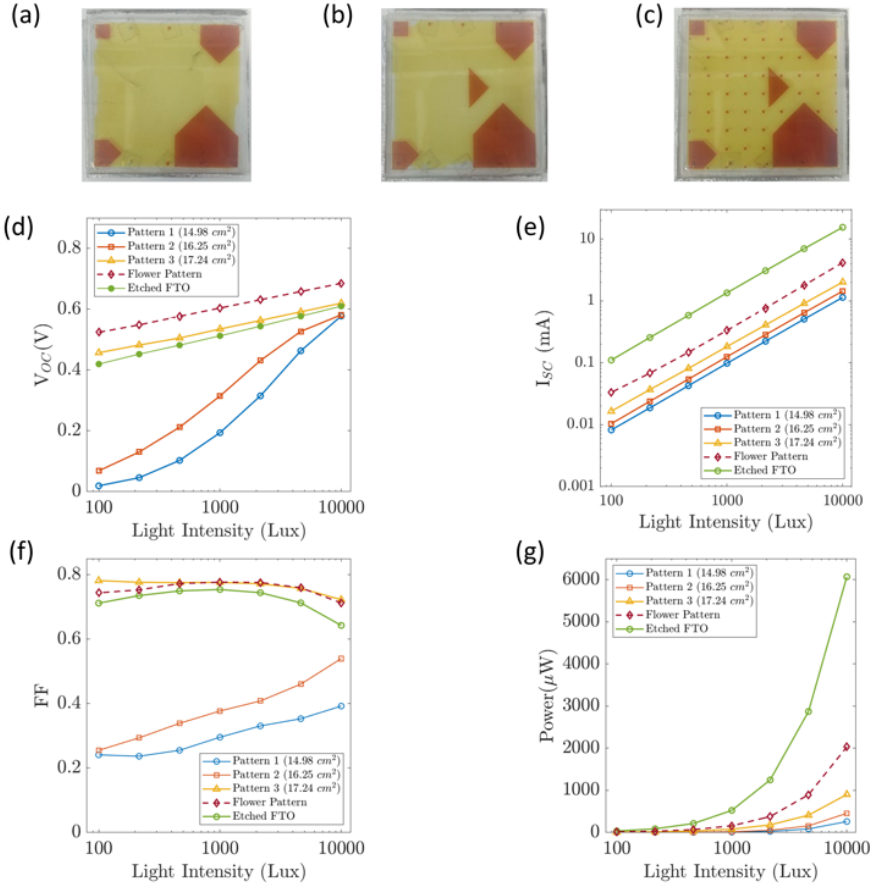

*Figure 6: (a) Pattern 1. (b) Pattern 2. (c) Pattern 3. (d) Open-circuit voltage output. (e) Short-circuit current output. (f) Fill factor output. (g) Absolute power output.*

**Table 3.** Confusion Matrix of best evaluated model from pattern 3 (17.24 cm$^2$)

|  | Right | Left | Up | Down | Up-Right | Up-Left | Down-Right | Down-Left |
|---|---|---|---|---|---|---|---|---|
| Right | 116 | 1 | 0 | 0 | 2 | 1 | 0 | 0 |
| Left | 0 | 116 | 0 | 0 | 0 | 3 | 1 | 0 |
| Up | 0 | 0 | 95 | 6 | 2 | 14 | 3 | 0 |
| Down | 0 | 0 | 13 | 83 | 13 | 2 | 2 | 7 |
| Up-Right | 0 | 4 | 1 | 0 | 104 | 6 | 5 | 0 |
| Up-Left | 0 | 2 | 36 | 1 | 1 | 75 | 5 | 0 |
| Down-Right | 2 | 0 | 0 | 0 | 0 | 1 | 114 | 3 |
| Down-Left | 0 | 0 | 4 | 18 | 0 | 2 | 0 | 96 |

True Class (rows) / Predicted Class (columns)

## 4. CONCLUSION

Some applications involving IoT nodes require several sensors, actuators and displays to provide its functionality, increasing the energy consumption demand of the IoT node. This paper investigates a method of integrating a monolithic DSSCs to not only harvest energy but also self-power its interactive features. We demonstrated that by monitoring the photocurrent output of an asymmetrically patterned monolithic (i.e., single cell) DSSC, and using machine learning, it can recognise simple hand gestures (four linear directions), achieving an accuracy prediction of 97.71 %. Compared with previous self-powered hand gesture recognition research, this method provides a simpler and visually appealing UI, as opposed to attaching several individual photodiodes/PV cells to the system [7][8][9]. We also optimised the asymmetrical active area pattern of the DSSC through computational modelling, achieving an accuracy prediction of 83.23 % for eight different hand directional gestures. This DSSC pattern was also modified to demonstrate the capabilities of providing an aesthetically pleasing environment, alongside also detecting hand gestures. The alternative fabrication method of etching the FTO was also successful in detecting hand gestures. Compared to SolarGest by Ma et al. [11], being the only other previous research that achieved hand gesture recognition through a monolithic PV cell; this work provides several natural hand gestures recognised through a machine learning system that can also achieve live gesture recognition, not capable through using other methods. We have proven the concept of accurate gesture detection in patterned, monolithic solar cells as a means of simultaneous energy harvesting and control and interaction with some kind of IoT device. The work was carried out under controlled illumination conditions with a consistent light source incident on the PV cell. Future work will need to evaluate the accuracy of the gesture recognition in realistic scenarios which may require more sophisticated training models.

**Corresponding Author**
*Matthew Carnie - Materials Science and Engineering & SPECIFIC-IKC, Faculty of Science and Engineering, Swansea University, Swansea, UK
Email: m.j.carnie@swansea.ac.uk

**Author Contributions**
GT planned and caried out all the experimental work. AP helped with solar cell measurements created the model to determine most effective printed pattern. KS helped with electronics and signal processing. MC devised the project and initial ideas and helped plan the experimental work. All authors contributed to the preparation of the manuscript.
**Funding Sources**
All authors would like to thank EPSRC for funding the following projects: GENERATION (EP/W025396/1), ATIP (EP/T028513/1), SPECIFIC-IKC (EP/N020863/1) and PV Interfaces (EP/R032750/1)

**REFERENCES**

[1] "Internet of Things (IoT) connected devices installed base worldwide from 2015 to 2025," *Statista Research Department*, 2016. [Online]. Available: https://www.statista.com/statistics/471264/iot-number-of-connected-devices-worldwide/. [Accessed: 15-Apr-2020].

[2] P. Sparks, "The route to a trillion devices," *White Pap. ARM*, 2017.

[3] H. A. Khattak, M. A. Shah, S. Khan, I. Ali, and M. Imran, "Perception layer security in Internet of Things," *Futur. Gener. Comput. Syst.*, vol. 100, pp. 144–164, 2019.

[4] T. Sanislav, G. D. Mois, S. Zeadally, and S. C. Folea, "Energy Harvesting Techniques for Internet of Things (IoT)," *IEEE Access*, vol. 9, pp. 39530–39549, 2021.

[5] V. Pecunia, L. G. Occhipinti, and R. L. Z. Hoye, "Emerging indoor photovoltaic technologies for sustainable internet of things," *Adv. Energy Mater.*, vol. 11, no. 29, p. 2100698, 2021.

[6] M. Oudah, A. Al-Naji, and J. Chahl, "Hand Gesture Recognition Based on Computer Vision: A Review of Techniques," *J. Imaging*, vol. 6, no. 8, 2020.

[7] A. Mujahid *et al.*, "Real-Time Hand Gesture Recognition Based on Deep Learning YOLOv3 Model," *Appl. Sci.*, vol. 11, no. 9, 2021.

[8] Y. Dong, J. Liu, and W. Yan, "Dynamic Hand Gesture Recognition Based on Signals From Specialized Data Glove and Deep Learning Algorithms," *IEEE Trans. Instrum. Meas.*, vol. 70, pp. 1–14, 2021.

[9] L. Guo, Z. Lu, and L. Yao, "Human-Machine Interaction Sensing Technology Based on Hand Gesture Recognition: A Review," *IEEE Trans. Human-Machine Syst.*, vol. 51, no. 4, pp. 300–309, 2021.

[10] D. Zhang *et al.*, "Flexible computational photodetectors for self-powered activity sensing," *npj Flex. Electron.*, vol. 6, no. 1, p. 7, 2022.

[11] D. Ma *et al.*, "SolarGest: Ubiquitous and Battery-Free Gesture Recognition Using Solar Cells," in *The 25th Annual International Conference on Mobile Computing and Networking*, 2019.

[12] A. Varshney, A. Soleiman, L. Mottola, and T. Voigt, "Battery-Free Visible Light Sensing," in *Proceedings of the 4th ACM Workshop on Visible Light Communication Systems*, 2017, pp. 3–8.

[13] Y. Li, T. Li, R. A. Patel, X.-D. Yang, and X. Zhou, "Self-Powered Gesture Recognition with Ambient Light," in *Proceedings of the 31st Annual ACM Symposium on User Interface Software and Technology*, 2018, pp. 595–608.

[14] H. Duan, M. Huang, Y. Yang, J. Hao, and L. Chen, "Ambient Light Based Hand Gesture Recognition Enabled by Recurrent Neural Network," *IEEE Access*, vol. 8, pp. 7303–7312, 2020.

[15] Y. K. Meena *et al.*, "PV-Tiles: Towards Closely-Coupled Photovoltaic and Digital Materials for Useful, Beautiful and Sustainable Interactive Surfaces," in *Proceedings of the 2020 CHI Conference on Human Factors in Computing Systems*, 2020, pp. 1–12.

[16] A. B. Muñoz-García *et al.*, "Dye-sensitized solar cells strike back," *Chem. Soc. Rev.*, vol. 50, no. 22, pp. 12450–12550, Nov. 2021.

[17] L. Xie *et al.*, "Recent progress of organic photovoltaics for indoor energy harvesting," *Nano Energy*, vol. 82, p. 105770, 2021.

[18]     Y. Cao, Y. Liu, S. M. Zakeeruddin, A. Hagfeldt, and M. Grätzel, "Direct contact of selective charge extraction layers enables high-efficiency molecular photovoltaics," *Joule*, vol. 2, no. 6, pp. 1108–1117, 2018.

[19]     J. Tang, S. Alelyani, and H. Liu, "Feature selection for classification: A review," *Data Classif. Algorithms Appl.*, p. 37, 2014.

**Supplementary Information for:**

Simultaneous Energy Harvesting and Hand Gesture Recognition in Large Area Monolithic Dye-Sensitized Solar Cells

Gethin Thomas, Adam Pockett, Kris Seunarine and Matt Carnie*

Materials Science and Engineering & SPECIFIC-IKC, Faculty of Science and Engineering, Swansea University, Swansea, UK

*m.j.carnie@swansea.ac.uk

## 1. THRESHOLDING TECHNIQUE

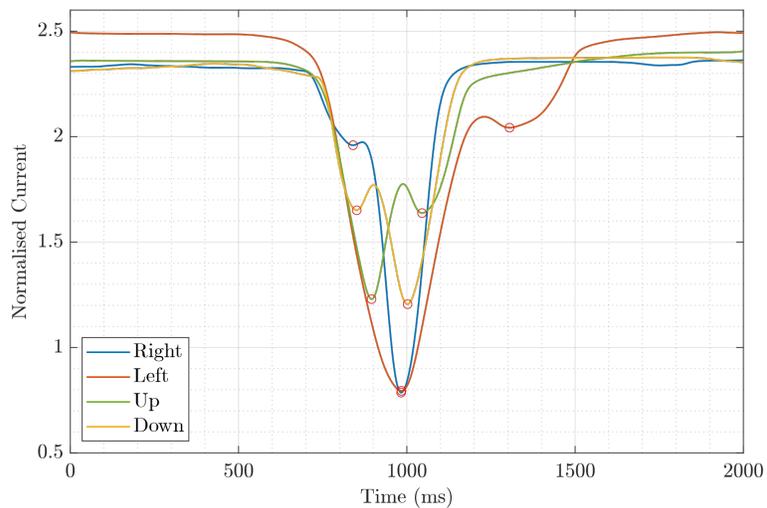

**Figure S1:** First Trough detection for each gesture photocurrent

**Table S1.1: Confusion Matrix Using Threshold Detection**

| True Class | Right | Left | Up | Down |
|---|---|---|---|---|
| Right | 0 | 14 | 226 | 0 |
| Left | 2 | 3 | 200 | 35 |
| Up | 17 | 97 | 103 | 23 |
| Down | 0 | 128 | 112 | 0 |

Predicted Class

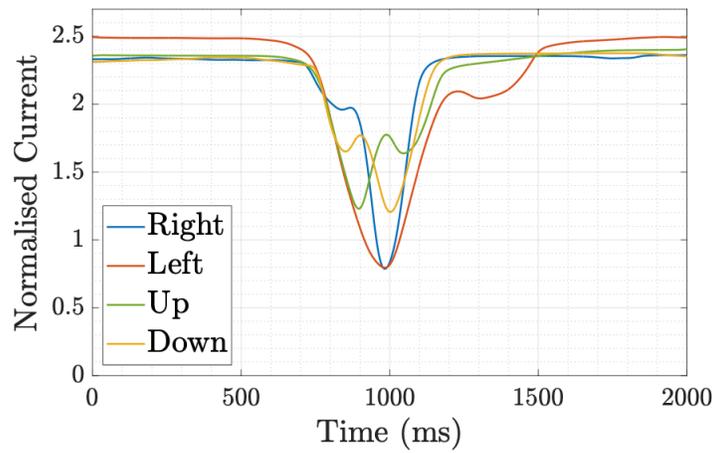

(a) Over 2500 Lux

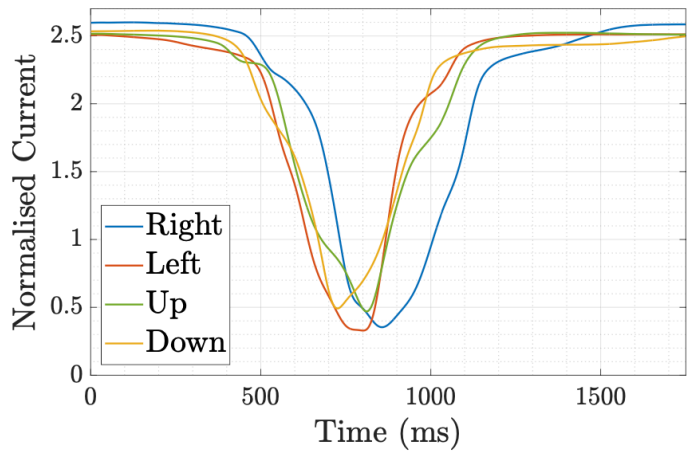

(b) Under 1000 Lux

**Figure S2:** Comparing photocurrent signals under (a) over 2500 lux and (b) under 1000 lux

**2. PATTERN DESIGN AND PHOTOCURRENT RESPONSE**

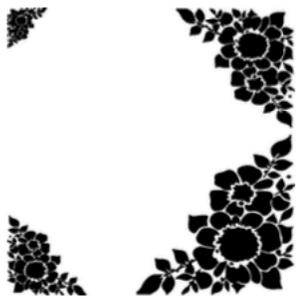 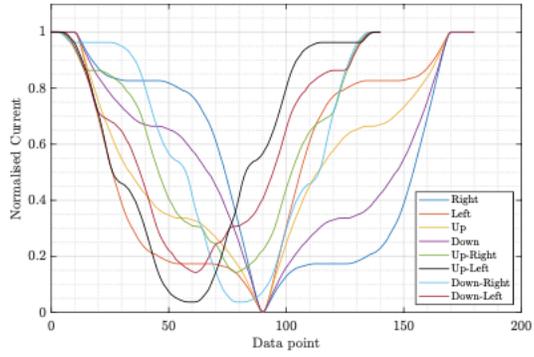

**Figure S3:** Design 1

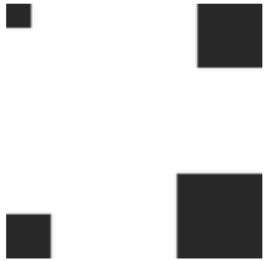 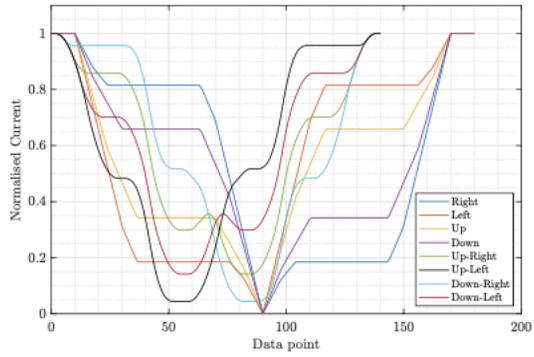

**Figure S4:** Design 2

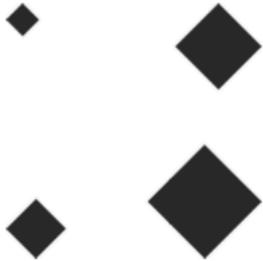 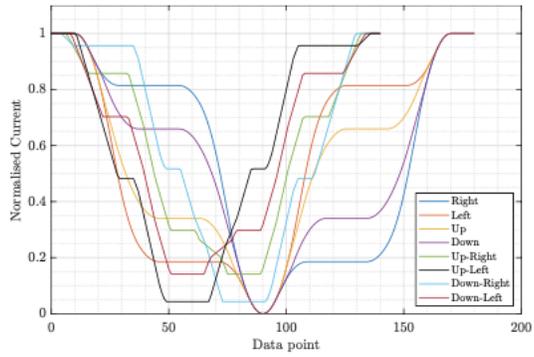

**Figure S5:** Design 3

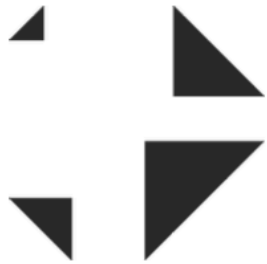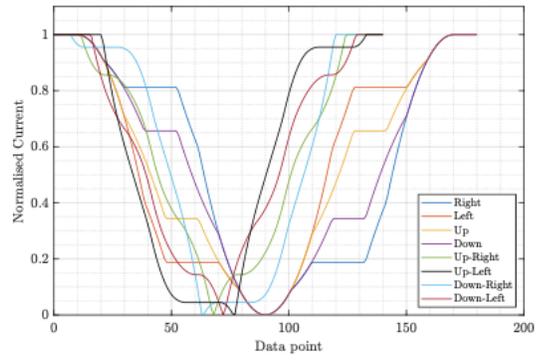

**Figure S6:** Design 4

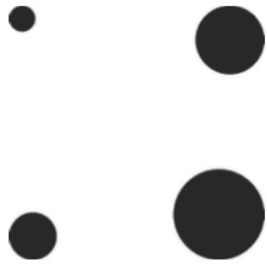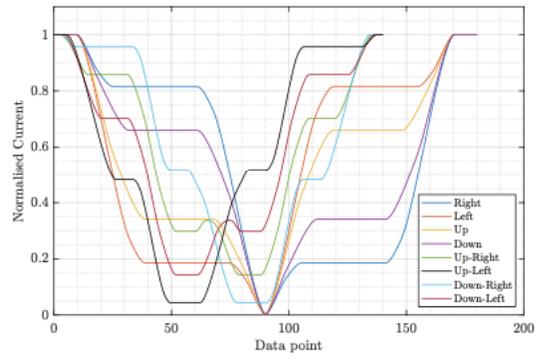

**Figure S7:** Design 5

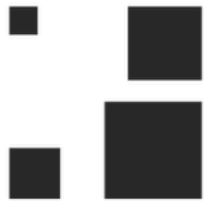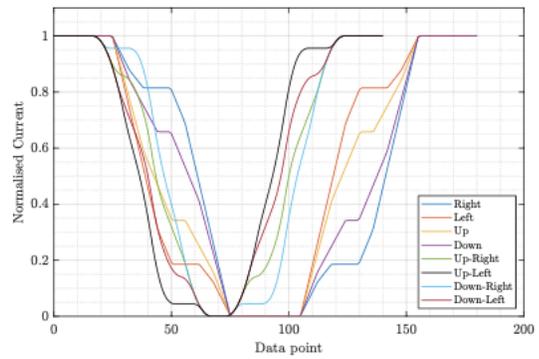

**Figure S8:** Design 6

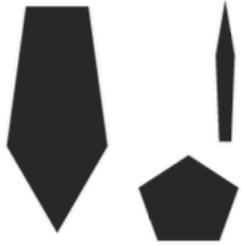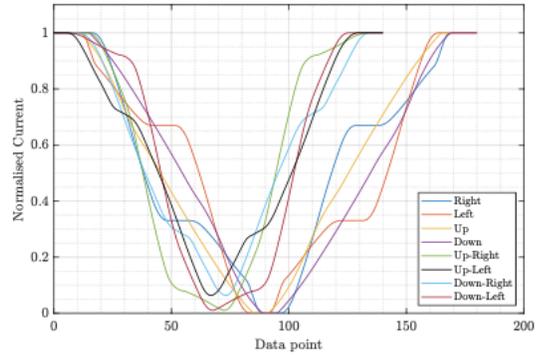

**Figure S9:** Design 7

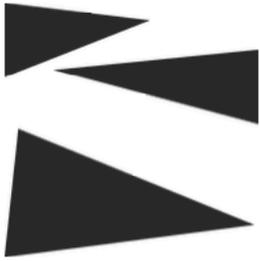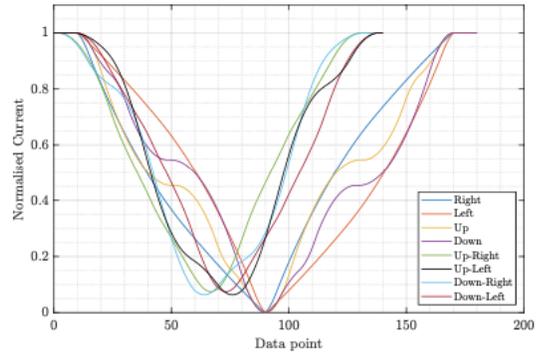

**Figure S10:** Design 8

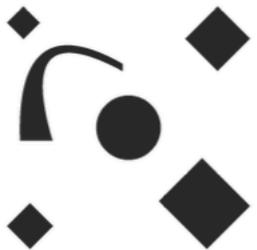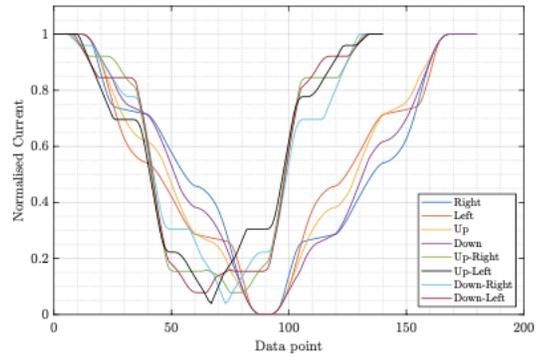

**Figure S11:** Design 9

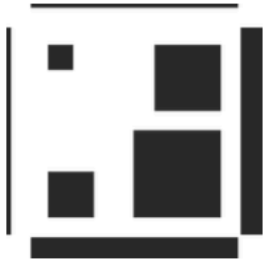 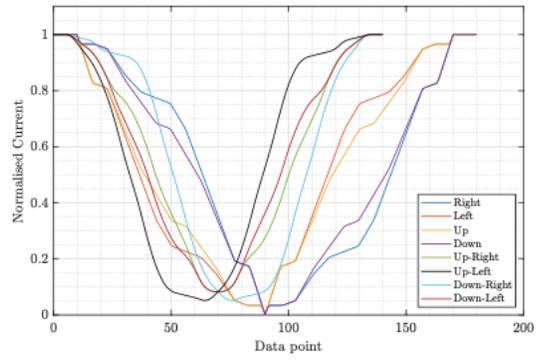

**Figure S12:** Design 10

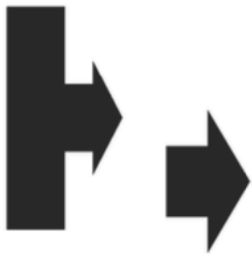 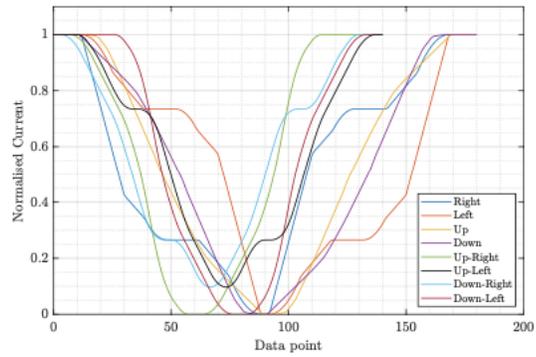

**Figure S13:** Design 11

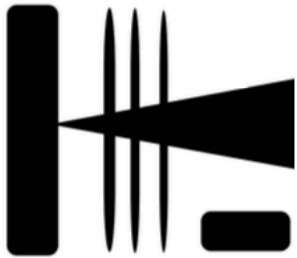 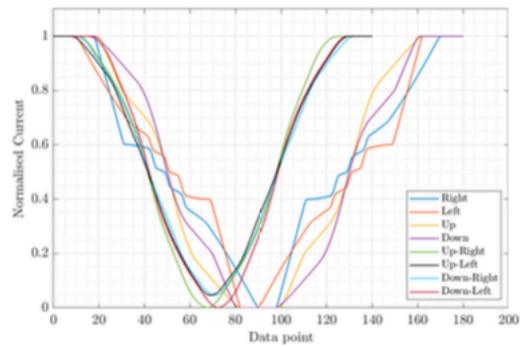

**Figure S14:** Design 12

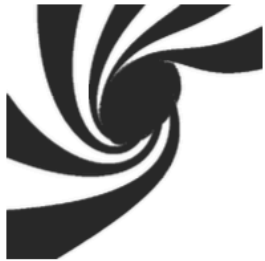 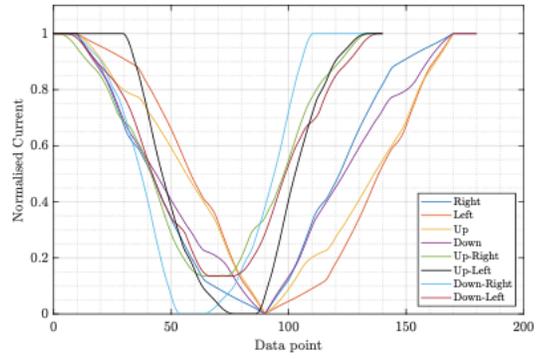

**Figure S15:** Design 13

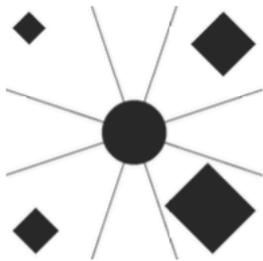 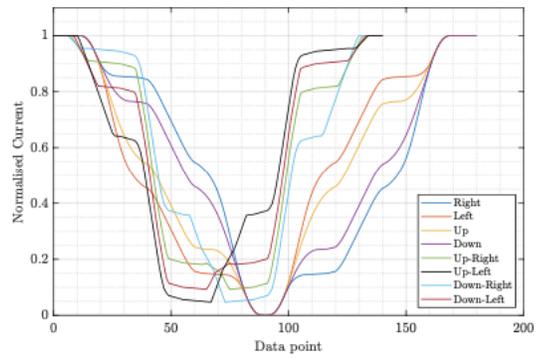

**Figure S16:** Design 14

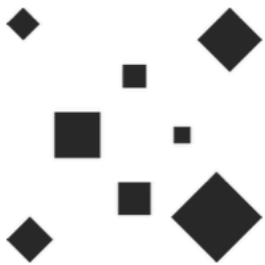 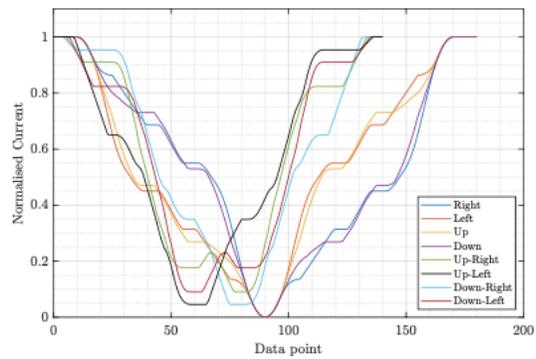

**Figure S17:** Design 15

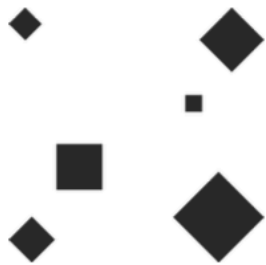 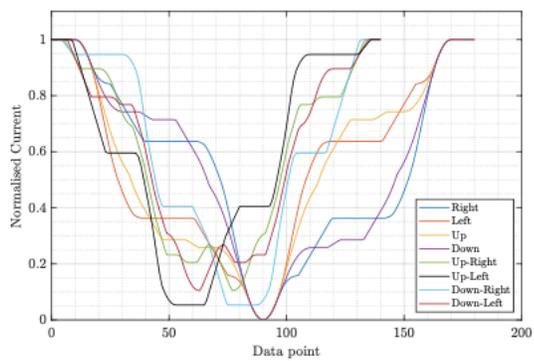

**Figure S18:** Design 16

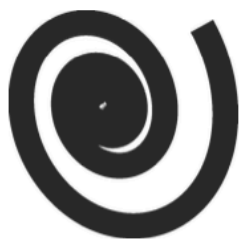 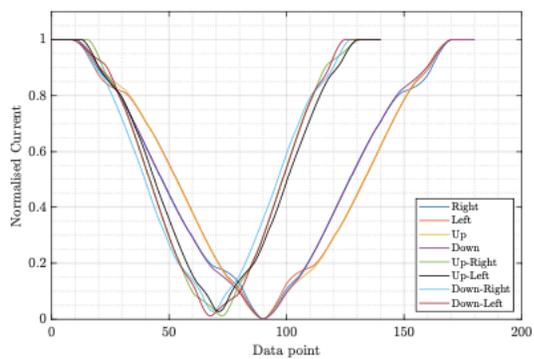

**Figure S19:** Design 17

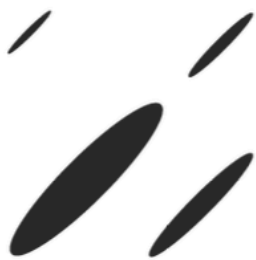 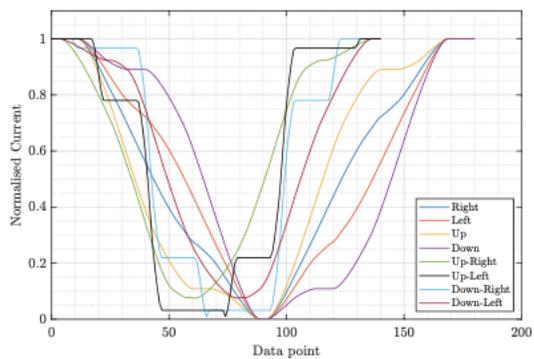

**Figure S20:** Design 18

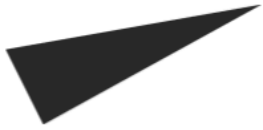
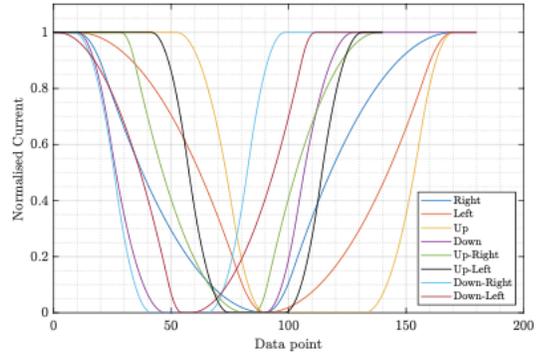

**Figure S21:** Design 19

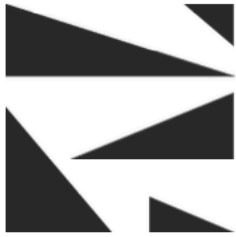
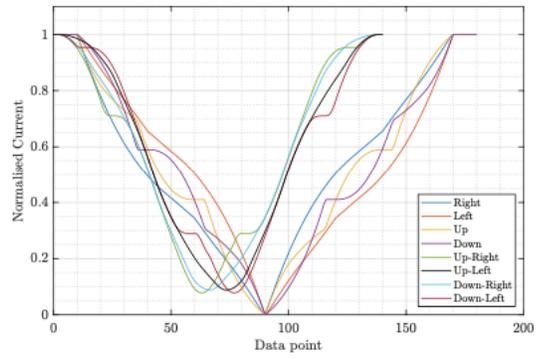

**Figure S22:** Design 20

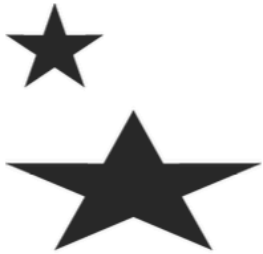
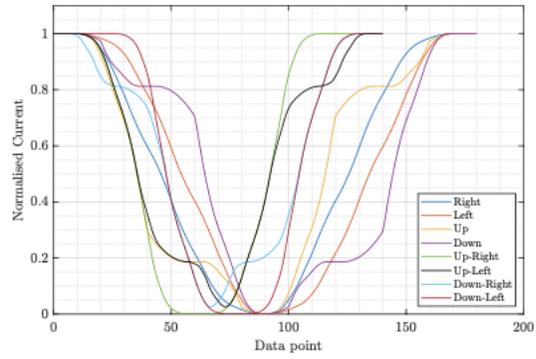

**Figure S23:** Design 21

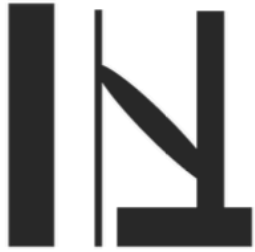 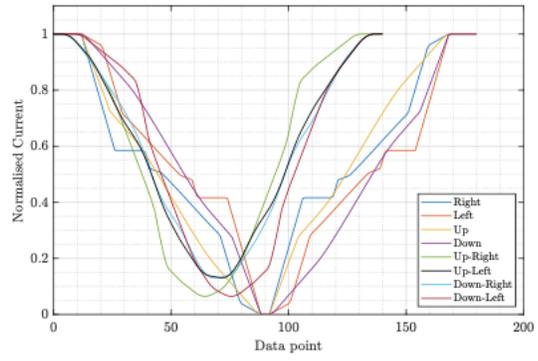

**Figure S24:** Design 22

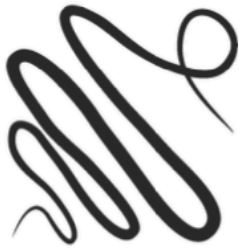 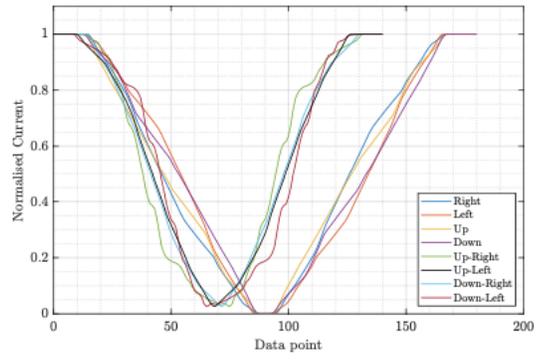

**Figure S25:** Design 23

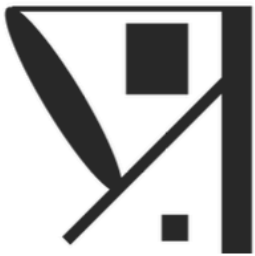 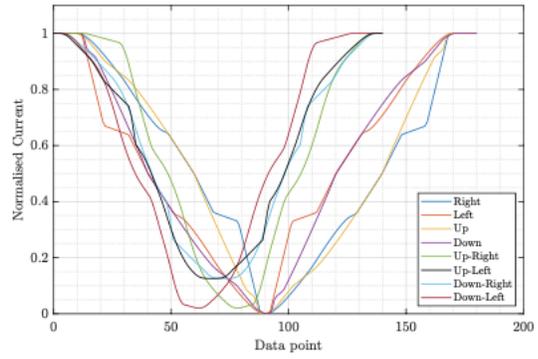

**Figure S26:** Design 24

## 3. OPTIMISATION

**Step 1.**

By increasing the area of the largest circle from design 5, the signal differences between up-left and down-right improves, however, this reduces the curve feature.

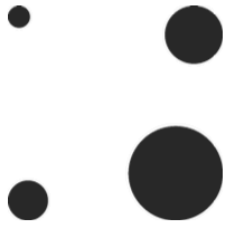 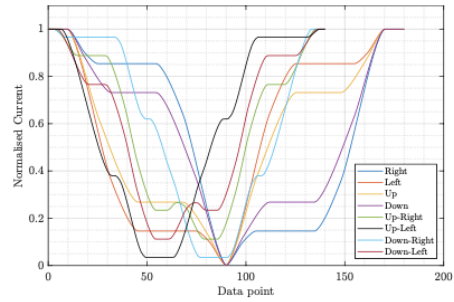

**Figure S27:** Step 1

**Step 2.**
By shifting the two medium area circles towards the smallest, this increases the 0 gradient for up-left and down-right, but now eliminates the distinct peak for up-right and down-left.

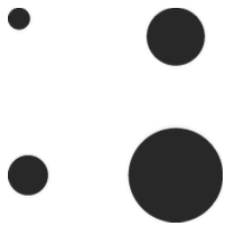 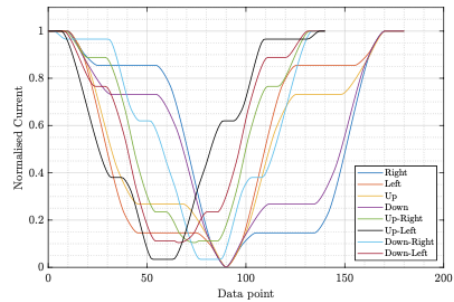

**Figure S28:** Step 2

**Step 3.**
The peaks are restored through squaring the circle edges.

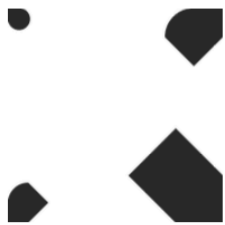 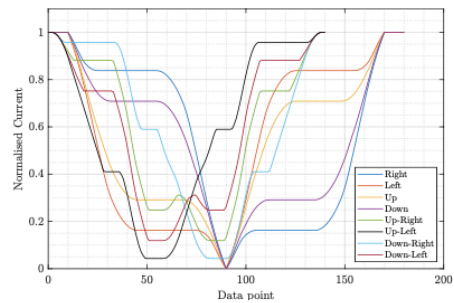

**Figure S29:** Step 3

**Step 4.**
Increasing the overall area. Weakens the signal slightly, but not enough to make a significant difference. The goal is to now make a larger difference signal between the pairs (down and right) and (left and up).

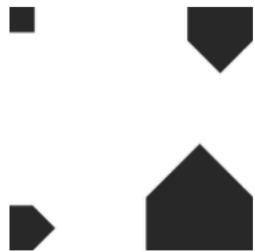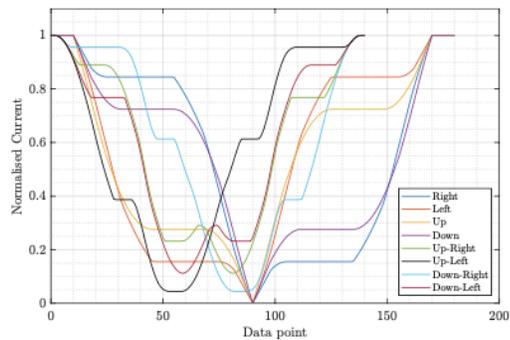

**Figure S30:** Step 4

### Step 5.
By squaring the edges of the two medium areas, this weakened the Up-Left and Down-right signal.

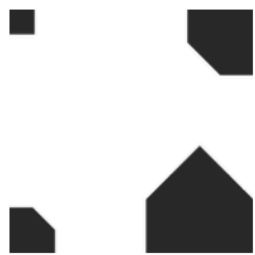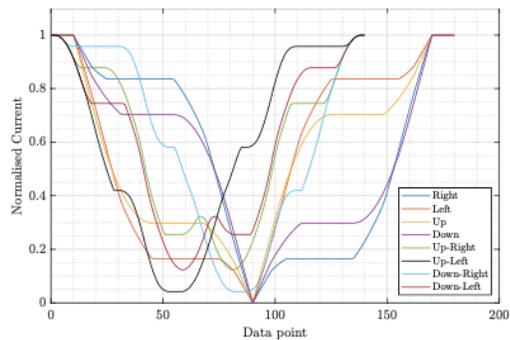

**Figure S31:** Step 5

### Step 6.
Altering pattern 4, increasing the top right corners shape area results in a larger difference between the two signals pairs mentioned in pattern 4, however has now weakened the peaks.

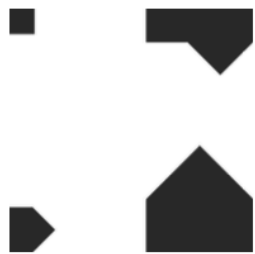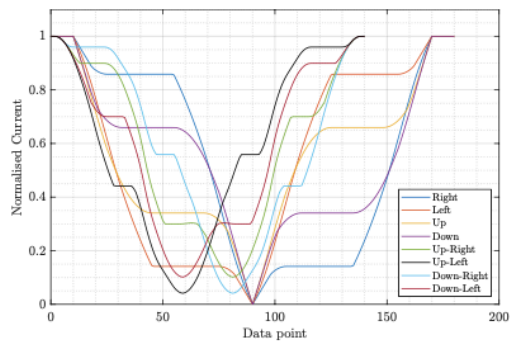

**Figure S32:** Step 6

### Step 7.
Increasing the area of the smallest square results in the lowest troughs of all diagonal gestures becoming similar, compared to pattern 4.

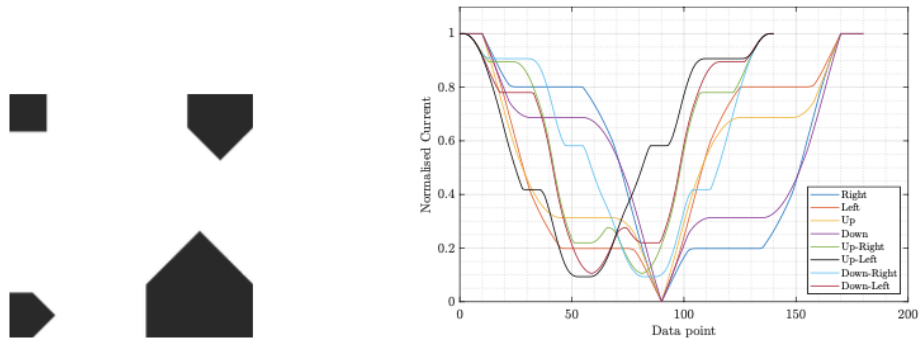

**Figure S33:** Step 7

### Step 8.
Step 4 seems to be the optimised pattern following design 5. This pattern now integrates part 2 for signal optimisation. The pattern below explains that the only possible location that the new shapes can appear is in the red circle box, otherwise, it can affect the shape of the signals previously optimised.

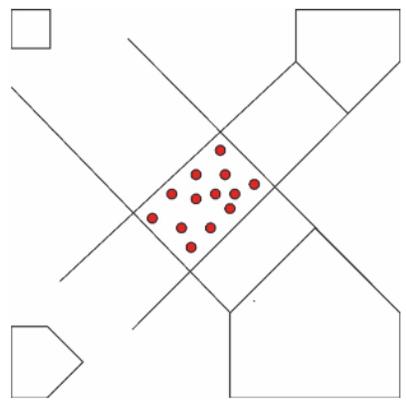

### Step 9.
By fully filling the area, it is seen that the up-left and down-right are too similar.

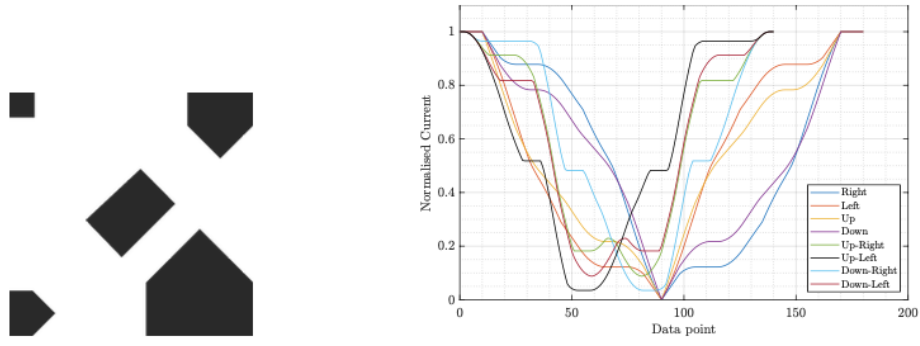

**Figure S34:** Step 9

### Step 10.
Breaking the centre rectangular into two parts of different sizes weakens the horizontal and vertical gesture signals.

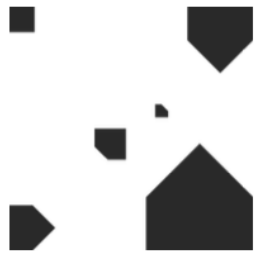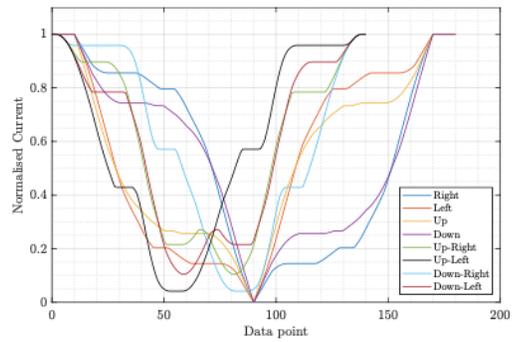

**Figure S35:** Step 10

**Step 11.**
The squares have a larger effect on the horizontal motion; however, it slightly weakens the vertical signal gestures.

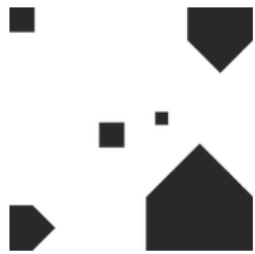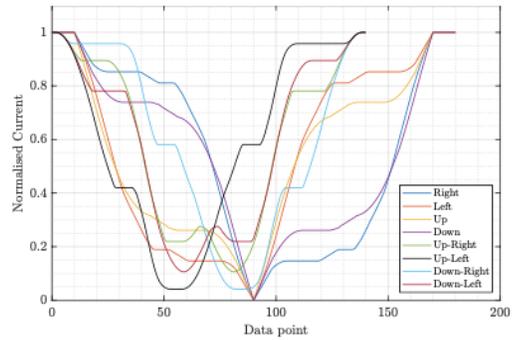

**Figure S36:** Step 11

**Step 12.**
The photocurrent signal gap of the left/up and down/right signal has decreased.

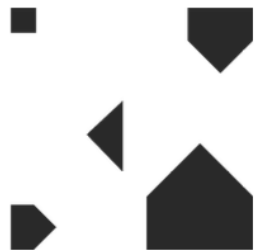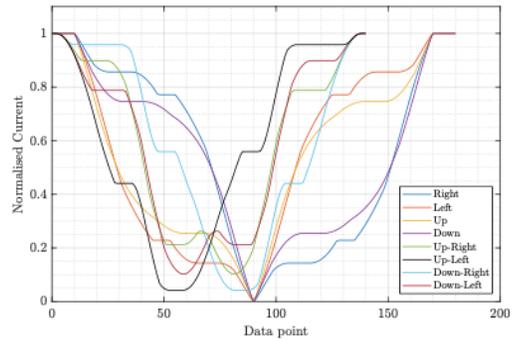

**Figure S37:** Step 12

**Step 13.**
This is an improvement from step 12, with a greater differential between the signals.

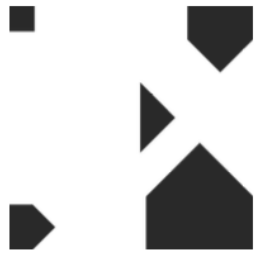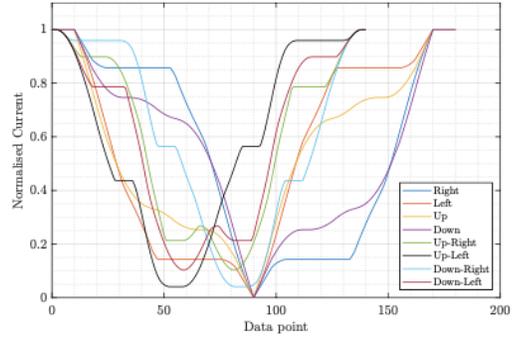

**Figure S38:** Step 13

### Step 14.
This decreases the right and left 0 gradient signal pattern, which becomes too similar to the up/down gesture.

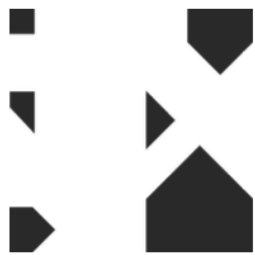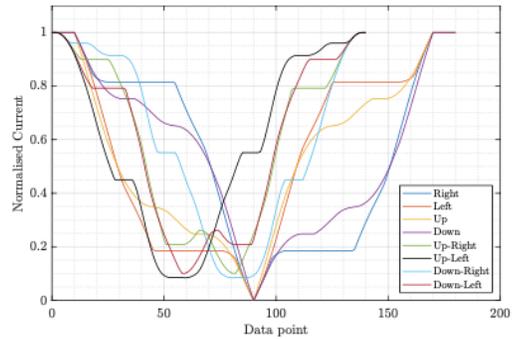

**Figure S39:** Step 14

### Step 15.
This again is a deterioration from the step 13 signal output. Overall, it is seen that step 13 is best following the addition of part 2.

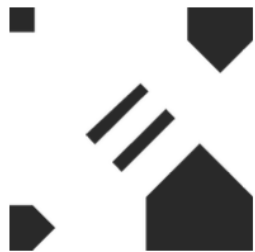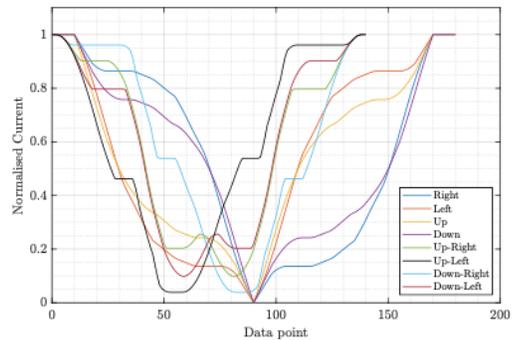

**Figure S40:** Step 15

### Step 16.
To further optimise the pattern design. A background can be added to slightly increase the active area of the solar cell. 1 mm diameter circles showed some promising response, however, the signal between up-left and down-right are too similar.

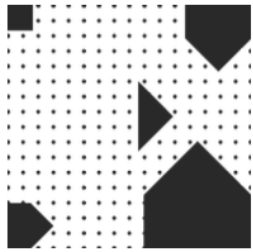
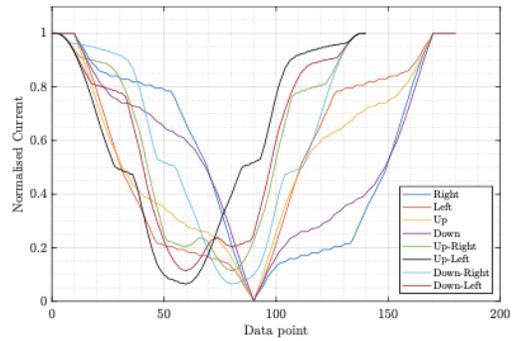

**Figure S41:** Step 16

## Step 17.
Spacing the 1 mm diameter circles resulted in a better deviation between the signals mentioned in step 16.

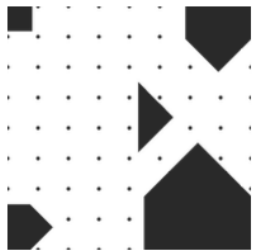
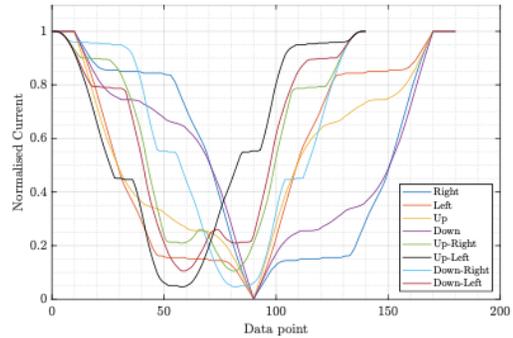

**Figure S42:** Step 17

## Step 18.
A larger diameter circle of 2 mm weakens the signal greatly.

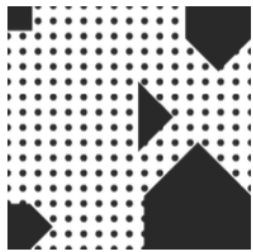
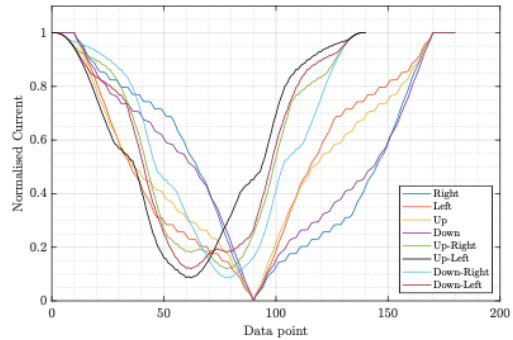

**Figure S43:** Step 18

## Step 19.
This is an improvement from step 18, however signals up-left and down-right are too similar.

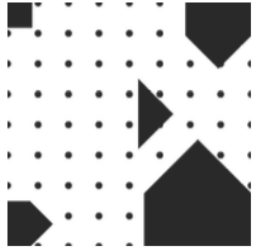
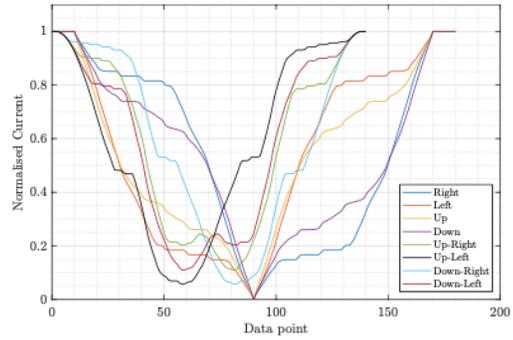

**Figure S44:** Step 19

## Step 20.

This is the best compromise between increasing the circles area and signal differential output. The circle diameter is 1.5 mm, with a good signal output like step 17.

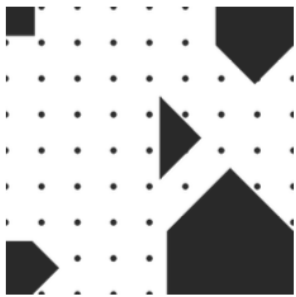
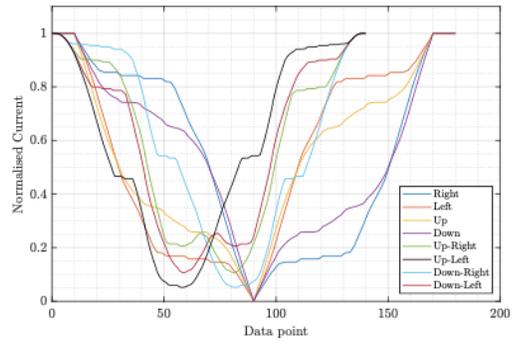

**Figure S45:** Step 20

**4. ELECTROLYTE VS VOID DISTANCE**

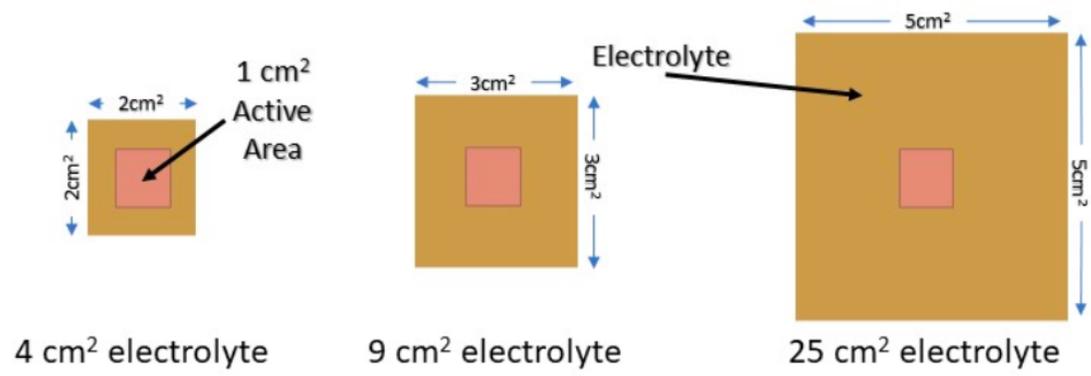

**Figure S46:** Changing the electrolyte area around a 1 $cm^2$ active area

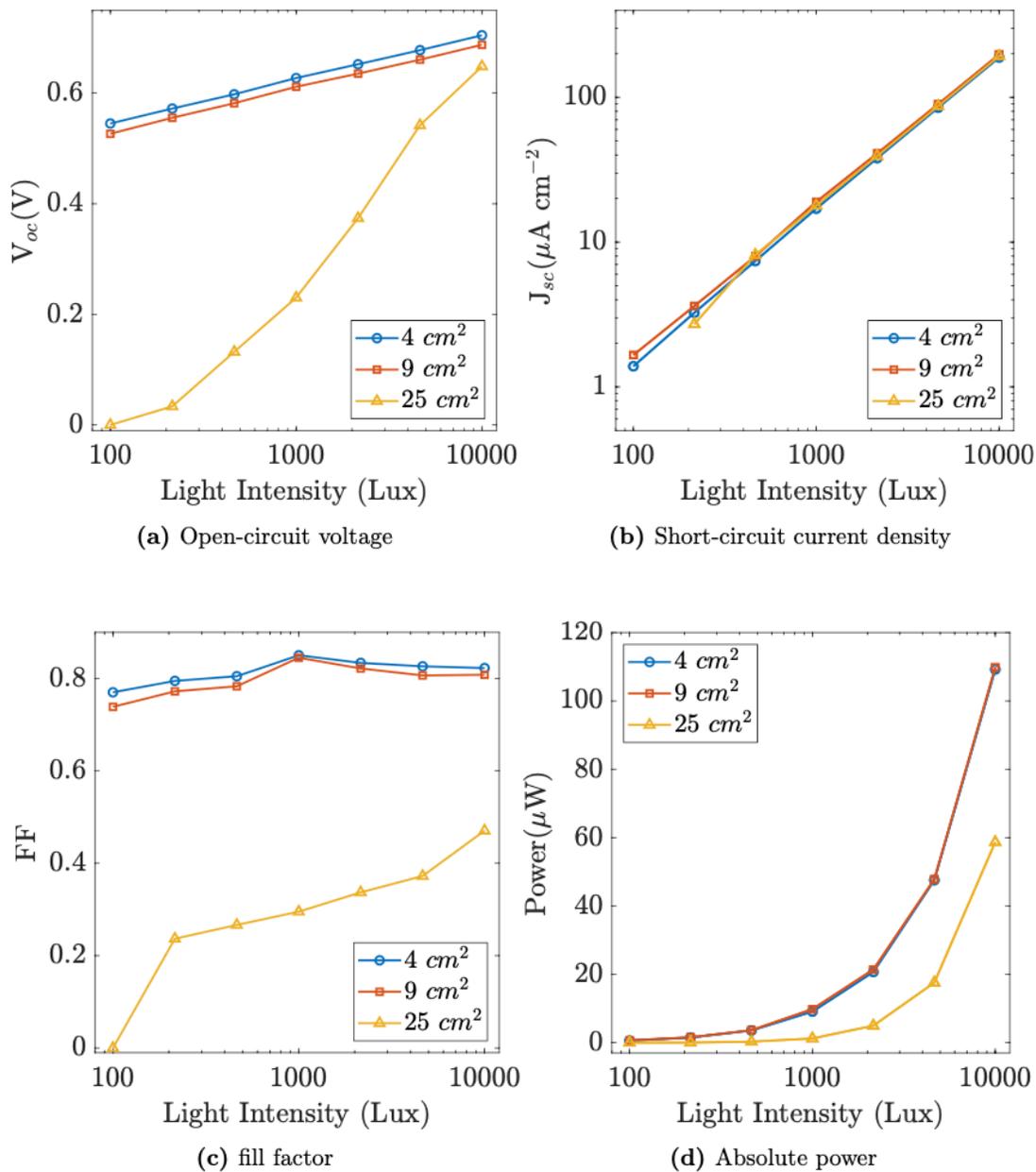

**Figure S47:** 50 mM (HM) DSSC performance characteristics of changing the electrolyte area around a 1 $cm^2$ active area